\RequirePackage{fix-cm}
\documentclass[onecollarge, natbib]{svjour2}
\bibpunct{[}{]}{,\!}{n}{}{,\!} 

\usepackage{amsmath}
\usepackage{amssymb}
\usepackage{bm}
\usepackage{color}
\usepackage{graphicx}
\usepackage{placeins}
\usepackage{supertabular}
\usepackage{xspace}
\usepackage{slashed}
\def\beq{\begin{equation}}
\def\eeq{\end{equation}}
\def\bea{\begin{eqnarray}}
\def\eea{\end{eqnarray}}
\def\nn{\nonumber}

\def\x{{\bm x}}
\def\y{{\bm y}}

\def\p{{\bm p}}

\def\ovl{\overline}
\def\sla{\slashed}
\def\widt{\widetilde}
\def\wS{\widt{S}}
\usepackage{mathptmx}      
\usepackage[mathscr,scaled=1.15]{urwchancal}
\DeclareFontFamily{OT1}{pzc}{}
\DeclareFontShape{OT1}{pzc}{m}{it}%
{<-> s * [1.15] pzcmi7t}{}
\DeclareMathAlphabet{\mathpzc}{OT1}{pzc}{m}{it}

\definecolor{purple}{rgb}{0.5,0,0.5}
\definecolor{blue}{rgb}{0.0,0,0.9}
\definecolor{prdblue}{rgb}{0.133,0.118,0.498}
\usepackage[colorlinks=true, pdfstartview=FitV, linkcolor=prdblue, citecolor= prdblue, 
urlcolor=prdblue]{hyperref}

\journalname{Few-Body Systems}
\begin{document}

\title{Quark propagator in Minkowski space
}

\titlerunning{Quark propagator in Minkowski space}

\author{E.~L.~Solis 
    \and
        C.~S.~R.~Costa
    \and
        V.~V.~Luiz
    \and
        G.~Krein
}


\institute{E.L. Solis, C.S.R. Costa, V.V. Luis,  G. Krein  \at
            Instituto de F\'{\i}sica Te\'orica, Universidade Estadual Paulista 
            \\ Rua Dr. Bento Teobaldo Ferraz, 271 - Bloco II, 01140-070 S\~ao Paulo, SP, Brazil.\\
              \\[-2ex]
            $\,$\\G. Krein [Corresponding author]\at
              \email{\href{mailto:gastao.krein@unesp.br}{gastao.krein@unesp.br}}\\
}

\date{Received:
28 January 2019
}

\maketitle

\begin{abstract}
The analytic structure of the quark propagator in Minkowski space is more complex than in 
Euclidean space due to the possible existence of poles and branch cuts at timelike momenta. 
These singularities impose enormous complications on the numerical treatment of the 
nonperturbative Dyson-Schwinger equation for the quark propagator. Here we discuss a 
computational method that avoids most of these complications. The method makes use of 
the spectral representation of the propagator and of its inverse. The use of spectral 
functions allows one to handle in exact
manner poles and branch cuts in momentum integrals. We obtain model-independent 
integral equations for the spectral functions and perform their renormalization by
employing a momentum-subtraction scheme. We discuss an algorithm for solving numerically the 
integral equations and present explicit calculations in a schematic model for the quark-gluon 
scattering kernel. 
\end{abstract}

\section{Introduction}
\label{intro}

Most of our understanding of the strong-coupling regime of quantum chromodynamics (QCD) comes from
studies employing mathematical methods formulated in Euclidean space. Lattice QCD is the prime 
example of a first-principles nonperturbative method formulated in Euclidean space{\textemdash}a 
review describing the theoretical foundations and calculation methods of lattice QCD can be found 
in section~17 of Ref.~\cite{Tanabashi:2018oca}, in which one also finds an extensive list of 
references and results of hadronic quantities calculated with this method. Another 
first-principles nonperturbative method is the functional approach in the continuum, founded 
on the functional renormalization group and the Dyson-Schwinger and Bethe-Salpeter-Faddeev equations. 
These equations can be formulated either in Euclidean space or in the Minkowski space, but it is the 
formulation in Euclidean space that this method achieves its most impressive 
successes{\textemdash}Refs.~\cite{Gies:2006wv,Cloet:2013jya,Aguilar:2015bud,Eichmann:2016yit,
Horn:2016rip,Fischer:2018sdj} are very recent reviews on results from this method. 

The  present paper is concerned with the continuum approach and 
is restricted to the Dyson-Schwinger equation for the quark propagator. We recall that the quark propagator 
is an essential ingredient in the Bethe-Salpeter-Faddeev equations to calculate e.g. bound-state properties 
like hadron masses, form factors, etc. Another motivation~\cite{Williams:1989tv,Krein:1990sf} for 
studying the quark propagator is that it gives insight on two of the most striking nonperturbative
phenomena in QCD, quark confinement and mass generation; the first 
is related to the notion that a single quark cannot propagate asymptotically, and the latter refers to 
the fact that hadrons like the proton and the neutron acquire their masses from almost massless quarks. 
In this framework, quark confinement is conjectured to be associated with dramatic changes in the analytic
structure of the propagator, and mass generation with a dramatic infrared enhancement of the momentum-dependent
quark-mass function. Possible changes in the analytic structure of the propagator include the absence of
a spectral representation respecting positivity constraints, presence of complex-mass poles and absence 
of a real mass-pole for timelike momenta. Much of what is presently known in this respect was gathered with
studies of the quark (and also gluon) propagator in Euclidean space, but very little is presently known
on the analytic structure of the propagator in Minkowski space, although a increasingly number of 
studies has appeared in the literature in the last years, a list of which is given in 
Refs.~\cite{Sauli:2001we,Sauli:2002tk,Sauli:2004bx,Sauli:2006ba,Biernat:2013ema,Sauli:2011xr,
Biernat:2013fka,Siringo:2016jrc,Biernat:2018tlj,Siringo:2017ide}. One reason for this disparity in favor of the
Euclidean formulation is that the analytic structure of the quark propagator (and of any other quark-gluon 
correlation function) in Minkowski space is more complicated than in Euclidean space. This is due to the 
possible existence of poles and branch cuts at timelike momenta, features that impose enormous complications 
on the numerical treatment of the nonperturbative Dyson-Schwinger equation (DSE) for the propagator. The
same sort of complications appear in the Bethe-Salpeter equation (BSE) formulated in Minkoswki 
space~\cite{Wick:1954eu,Itzykson:1980rh}{\textemdash}a partial list of references on recent
studies of the Minkowski-space BSE can be found in Refs.~\cite{Karmanov:2005nv,Sauli:2008bn,
Carbonell:2010zw,Sauli:2012xj,Carbonell:2014dwa,Gutierrez:2016ixt,dePaula:2016oct,Carbonell:2017isq,
Ydrefors:2019jvu}.

In this paper, we present a computational method for solving the Minkowski-space DSE that avoids most 
of these complications. The method makes use of the spectral representation of the propagator and of its 
inverse. Instead of solving the DSE for the momentum-dependent mass and wave-function 
renormalization functions, a common practice in the Euclidean formulation, in the spectral 
method one solves for the spectral functions of the propagator. One of the advantages of this method 
is that when performing the momentum integrals in the DSE, one does not need to know beforehand 
the singularities in the quark propagator, only those in the quark-gluon kernel (comprised by the
product of the gluon propagator and quark-gluon vertex function). That is, poles and branch cuts in 
the scattering kernel can be handled in an exact manner. The method has been used in the past in 
the context of 
meson-baryon effective Lagrangians~\cite{Brown:1968wi,Krein:1993jb,Bracco:1993cy,daRocha:1997vk}, 
but it needs adjustments to peculiarities of QCD. One of the most important adjustments, which is 
discussed in this paper, is the renormalization procedure. Spectral representations have been used 
recently to solve the DSE for QCD in Refs.~\cite{Sauli:2004bx,Sauli:2011xr,Tripolt:2018qvi,
Wang:2018osm,Duarte:2019pr} and for QED in Ref.~\cite{Sauli:2002tk,Jia:2017niz}.

We describe the general formulation of the spectral representation of the propagator in the next
section, in which we also discuss the renormalization scheme. We show how the DSE
for the quark propagator can be cast in a form convenient for solving it numerically in terms of
its spectral functions. Although our primary aim in this paper is to discuss the formalism, we 
illustrate its use in a schematic model for the quark-gluon kernel. This illustration is
presented in section~\ref{sec:model} including explicit numerical solutions for a selected set of
parameters. Although the kernel we use is a very crude representation of the singularity structure 
one can expect from a more realist Minkowski-space quark-gluon kernel, it nevertheless reveals 
illuminating model-independent features of the solutions, such as the presence of poles and branch
cuts, and positivity violation. Our conclusions and perspectives for realistic applications to hadron
structure are presented in section~\ref{sec:concl}.

%
\section{Spectral representation}
\label{sec:spectral}

In this section we discuss the spectral representation of the propagator of a 
spin-1/2 fermion in Minkowski space. The discussion is conducted in a model-independent
way. Issues related to positivity violation in the spectral functions and existence
of complex-mass poles are postponed to the next section. 
We start with a standard textbook discussion of the spectral representation of the propagator 
of a spin-1/2 fermion in terms of two spectral functions having support on the positive real 
axis. Then we show that the two spectral functions can be replaced by a single spectral 
function which has support over the entire real axis. Next, we project out the Dirac-matrix 
structure of the propagator to obtain a spectral representation in terms of a scalar function 
that depends on a complex-energy variable. Finally, we discuss the renormalization of the 
propagator.

\subsection{General properties}
\label{subsec:general}

Let $\psi_\Lambda(x)$ and $m_\Lambda$ be respectively the fermion bare field operator 
and mass. Here, $\Lambda$ indicates a regulatization cutoff. The bare field and mass
are related to the renormalized field $\psi$ and mass $m$ by renormalization factors~as
\beq
\psi_\Lambda(x) = \sqrt{Z_\psi} \, \psi(x), \hspace{1.0cm}
m_\Lambda = Z_m \, m.
\label{ren-const}
\eeq
The renormalized propagator is defined~by  
\bea
i S_{\alpha \beta} (x - y) =   \langle \Omega | T [\psi_{\alpha}(x) 
{\ovl\psi}_{\beta}(y)]| \Omega \rangle 
= Z^{-1}_\psi \, i S_{\Lambda\,\! \alpha \beta} (x - y)  , 
\label{def-prop}
\eea
where $|\Omega\rangle$ is the vacuum state, and $\alpha,\beta$ are Dirac
indices; internal-symmetry indices are suppressed for simplicity. The Fourier transform 
$S_{\alpha\beta}(p)$ of $S_{\alpha \beta} (x - y)$ is defined by 
\beq
S_{\alpha \beta} (x - y) = \int \frac{d^4 p}{(2\pi)^4} \, e^{ - i p\cdot (x-y)} 
\, S_{\alpha\beta}(p).
\eeq
We have not made explicit in $Z_\psi$, $Z_m$ and $m$ the dependence on the 
renormalization scale~$\mu$; this dependence will be inserted when strictly
needed.

When parity is a good quantum number, $S_\Lambda(p)$ can be written as (suppressing 
the Dirac indices)
\bea
S_\Lambda(p) &=& \frac{1}{A_\Lambda(p^2) \sla{p} - B_\Lambda(p^2) + i \epsilon} 
= \frac{1}{A_\Lambda(p^2)} \, \frac{1}{\sla{p} - M_\Lambda(p^2) + i \epsilon} \nn \\
&=& Z_\psi\,S(p) = Z_\psi\, \frac{1}{A(p^2) \sla{p} - B(p^2) + i \epsilon} \nn \\
&=& \frac{Z_\psi}{A(p^2)} \, \frac{1}{\sla{p} - M (p^2) + i \epsilon},
\label{S-AB}
\eea
where $A_\Lambda(p^2)$, $B_\Lambda(p^2)$, and $M_\Lambda(p^2) = B_\Lambda(p^2)/A_\Lambda(p^2)$
are Lorentz scalar functions. The scalar functions $A(p^2)$ and
$B(p^2)$ of the renormalized propagator, $S(p)$, are related to the corresponding unrenormalized ones 
by $A(p^2) = Z_\psi \, A_\Lambda(p^2)$ and $B(p^2) = Z_\psi \,B_\Lambda(p^2)$, 
which imply $M(p^2) = M_\Lambda(p^2)$. 
The noninteracting propagator is given by 
\beq
S^{(0)}_\Lambda(p) = \frac{1}{\sla{p} - m_\Lambda + i \epsilon} 
= \frac{\slashed{p} + m_\Lambda}{p^2 - m^2_\Lambda + i \epsilon}.
\label{S0}
\eeq
The inverse of $S_{\Lambda}(p)$ is conveniently written as
\beq
S^{-1}_{\Lambda}(p) = [S^{(0)}_\Lambda(p)]^{-1} - \Sigma_{\Lambda}(p),
\label{S-inv}
\eeq
where $\Sigma_\Lambda(p) = \left(1 - A_\Lambda(p)\right)\sla{p} + \left(m_\Lambda + B_\Lambda(p)\right)$ 
is the self-energy. 

The bare propagator has a spectral representation~\cite{Itzykson:1980rh}:
\beq
S_\Lambda(p) = \int^{\infty}_0 d s^2 \, 
\frac{\rho_{1\,\!\Lambda}(s^2) \sla{p}  + \rho_{2\,\!\Lambda}(s^2)}
{p^2 - s^2 + i \epsilon},
\label{S-rho12}
\eeq
where the spectral functions $\rho_{1\,\!\Lambda}(s^2)$ and $\rho_{2\,\!\Lambda}(s^2)$ 
are real and satisfy the positivity conditions:
\beq
\rho_{1\,\!\Lambda}(s^2) \ge 0, \hspace{1.0cm} s \,\rho_{1\,\!\Lambda}(s^2) 
- \rho_{2\,\!\Lambda}(s^2) \ge 0,
\label{pos-rho12}
\eeq
where $s = + \sqrt{s^2}$. The derivation of this representation relies on the
general principles of quantum field theory of CPT and Lorentz invariance and 
unitarity of quantum mechanics. 

It is possible to work with a single spectral function,
instead of two, by enlarging the integration range to $-\infty$ to $+\infty$; namely:
\beq
S_\Lambda(p) = \int^{+\infty}_{-\infty} d\kappa \, \rho_\Lambda(\kappa) \,
\frac{ \sla{p} + \kappa}
{p^2 - \kappa^2 + i \epsilon},
\label{S-rho}
\eeq
where $\rho_{1\,\Lambda}$ and $\rho_2$ are related to $\rho$ through:
\beq
\rho_{\Lambda\,\!1}(\kappa^2) = \frac{\rho_\Lambda(\kappa) + \rho_\Lambda(-\kappa)}{2 \kappa}, 
\hspace{1.0cm}
\rho_{\Lambda\,\!2}(\kappa^2) = \frac{\rho_\Lambda(\kappa) - \rho_\Lambda(-\kappa)}{2} .
\label{rhos}
\eeq
The positivity conditions (\ref{pos-rho12}) imply:
\beq
\rho_\Lambda(\kappa) \ge 0.
\label{pos-rho}
\eeq
 
It is convenient to introduce the projection operators~\cite{Brown:1968wi,Krein:1993jb}
\beq
P_{\pm}(p) = \frac{1}{2} \left( 1 \pm \frac{\sla{p}}{w(p)} \right) \hspace{0.5cm}
\text{where} \hspace{0.5cm}
w(p) \equiv  
\begin{cases} 
\sqrt{p^2} = \sqrt{(p^{0})^2 - \p^2} ,       & p^2 > 0  \\[0.3true cm]
i {\sqrt{-p^2}} = i \sqrt{\p^2 - (p^{0})^2}, & p^2 < 0 .
\end{cases} 
\label{Ppm}
\eeq
They allow us to project out the Dirac structure of the propagator by writing 
(\ref{S-rho}) as
\beq
S_\Lambda(p) = P_+ (p) \, {\wS}_\Lambda(w(p) + i \epsilon) 
+ P_+ (p) \, {\wS}_\Lambda(- w(p) - i \epsilon),
\label{Stilde}  
\eeq
where ${\wS}_\Lambda(z)$ is the scalar function
\beq
{\wS}_\Lambda(z) = \int^{+\infty}_{-\infty} d\kappa \, \frac{\rho_\Lambda(\kappa)}{z - \kappa},
\label{spectr-Stil}
\eeq
with $z = \pm (w(p) + i \epsilon)$. The corresponding spectral representation for renormalized 
propagator, written in terms of a renormalized spectral function $\rho(\kappa)$, is given by
\beq
{\wS}(z) = \int^{+\infty}_{-\infty} d\kappa \, \frac{\rho(\kappa)}{z - \kappa},
\label{wS-ren}
\eeq 
where, due to (\ref{def-prop}), $\rho_\Lambda(\kappa)$ and $\rho(\kappa)$ are related by the
field renormalization constant $Z_\psi$ as
\beq
\rho_\Lambda(\kappa) = Z_\psi\,\rho(\kappa).
\label{rel-rhos}
\eeq
As mentioned above, $Z_\psi = Z_\psi(\mu)$, where $\mu$ is the renormalization scale. Therefore, 
$\rho(\kappa) = \rho(\kappa,\mu)$, since $\rho_\Lambda(\kappa)$ is independent of~$\mu$. In addition, 
from (\ref{wS-ren}) one has that 
\beq
\rho(\kappa) = -\frac{1}{2\pi i} \left[{\wS}(\kappa+i\epsilon) - {\wS}(\kappa - i\epsilon)\right].
\label{rho-Stil}
\eeq

A spectral representation analogous to the one in (\ref{S-rho12}) exists for the vacuum 
expectation value of the anticommutador 
$\{\psi_{\Lambda\,\!\alpha}(x),{\ovl\psi}_{\Lambda\,\!\beta}(y)\}$~\cite{Itzykson:1980rh}:
\beq
\langle \Omega|\{\psi_{\Lambda\,\!\alpha}(x),{\ovl\psi}_{\Lambda\,\!\beta}(y)\}|\Omega\rangle
= i \int^{+\infty}_{-\infty} d\kappa\, \rho_\Lambda(\kappa) \left(i \sla\partial_x 
+ \kappa\right)_{\alpha\beta}
\Delta(x-y;\kappa),
\label{spect-anti}
\eeq
where $\rho_\Lambda(\kappa)$ is the same spectral function appearing in (\ref{S-rho}) and 
\beq
\Delta(x-y;\kappa) = i \int \frac{d^3p}{2p^0(2\pi)^3} \left[ e^{-ip\cdot(x-y)} 
- e^{ip\cdot(x-y)}\right],
\label{Delta-k}
\eeq
with $p^0 = \sqrt{\p^2 + \kappa^2}$. For $x^0 = y^0$, the left hand side of (\ref{spect-anti})
gives the equal-time anticommutador:
\beq
\{\psi_{\Lambda\,\!\alpha}(x^0, x),{\ovl\psi}_{\Lambda\,\!\beta}(y^0,\y)\}_{x^0 = y^{0}}
= i \delta^{(3)}(\x - \y) (\gamma^0)_{\alpha\beta}.
\eeq
Using that $\Delta(x-y;\kappa)|_{x^0=y^0}=0$ and $[\partial_0\Delta(x-y;\kappa)]_{x^0=y^0}
=-\delta^{(3)}(\x - \y)$, one obtains the important result
\beq
\int^{+\infty}_{-\infty}d\kappa\,\rho_\Lambda(\kappa) = 1.
\label{int-rho}
\eeq

This result is important for two main reasons. First, it allows us to relate the renormalization 
constant $Z_\psi$ to the renormalized spectral function $\rho(\kappa)$ by using (\ref{rel-rhos}):
\beq
Z^{-1}_{\psi} = \int^{+\infty}_{-\infty}d\kappa\,\rho(\kappa) .
\label{Z-rho}
\eeq
We will come back to this relationship in section~\ref{sec:model}, in which we discuss numerical 
results. Second, it implies that ${\wS}_\Lambda(z)$ has no zeros or poles off the real axis 
for positive $\rho_\Lambda(\kappa)$. This is shown as follows. Taking $z = x + iy$, with 
$x$ and $y$ real, one has 
\beq
{\wS}_\Lambda(z) = {\wS}_\Lambda(x+iy) = \int^{+\infty}_{-\infty} d\kappa \, 
\frac{\rho_\Lambda(\kappa)}{x + iy - \kappa} 
= \left(x - i y\right) \int^{+\infty}_{-\infty} d\kappa \, 
\frac{\rho_\Lambda(\kappa)}{(x-\kappa)^2 + y^2}.
\eeq
The imaginary part of ${\wS}_\Lambda(z)$ is then given by
\beq
{\rm Im} \, {\wS}_\Lambda(z) = - y  \int^{+\infty}_{-\infty} d\kappa \, 
\frac{\rho_\Lambda(\kappa)}{(x-\kappa)^2 + y^2} .
\eeq 
This is zero if $y = 0$ when $\rho_\Lambda(\kappa) > 0$; therefore, if there is
a zero in ${\wS}_\Lambda(z)$, it must lie on the real axis. When $\rho_\Lambda(\kappa)$
is not positive, one can have singularities off the real axis.

The absence of a zero off the real axis in ${\wS}_\Lambda(z)$ means that the inverse of 
the propagator does not have a pole off the real axis. This feature allows us to write a 
a spectral representation for the inverse of ${\wS}_\Lambda(z)$:
\beq
{\wS}^{-1}_\Lambda(z) = z - m_\Lambda 
-  \int^{+\infty}_{-\infty} d\kappa \, 
\frac{\sigma_\Lambda(\kappa)}{z-\kappa},
\label{Sig-tilde}
\eeq
and $S^{-1}_\Lambda(p)$ is obtained from this expression making use of the 
projection operators $P_{\pm}(p)$:
\beq
S^{-1}_\Lambda(p^2) = P_+(p) \, {\wS}^{-1}(w(p)+ i\epsilon) 
+ P_-(p) \, {\wS}^{-1}(-w(p) - i\epsilon).
\label{S-1-proj}
\eeq
One can easily show that ${\wS}^{-1}_\Lambda(z)$ also does not have zeros off the real 
axis if $\sigma_\Lambda(\kappa) > 0$; in this case, ${\wS}_\Lambda(z)$ has no poles off the 
real~axis. The renormalized inverse is 
\beq
{\wS}^{-1}(z) = Z_\psi\,  {\wS}^{-1}_\Lambda(z) = Z_\psi (z - Z_m m)
-  \int^{+\infty}_{-\infty} d\kappa \, 
\frac{\sigma(\kappa)}{z-\kappa}.
\label{Sig-ren}
\eeq
where the renormalized spectral function $\sigma(\kappa)$ is given by
\beq
\sigma_\Lambda(\kappa) = Z^{-1}_\psi\,\sigma(\kappa).
\label{ren-sig}
\eeq
Again, we have suppressed the $\mu$ dependence the spectral function.

From (\ref{Sig-ren}) one has that 
\beq
\sigma(\kappa) = \frac{1}{2\pi i} \left[{\wS}^{-1}(\kappa+i\epsilon) - {\wS}^{-1}(\kappa - i\epsilon)\right].
\label{sig-S-1til}
\eeq  
One can obtain a relationship
between the $\sigma(\kappa)$ and $\rho(\kappa)$ spectral functions: using the identity
\beq
{\wS}^{-1}(\kappa + i\epsilon) - {\wS}^{-1}(\kappa-i\epsilon) 
= {\wS}^{-1}(\kappa + i\epsilon) {\wS}^{-1}(\kappa-i\epsilon) 
\left[{\wS}(\kappa - i\epsilon) - {\wS}(\kappa+ i\epsilon)  \right],
\eeq
and taking into account (\ref{sig-S-1til}) and (\ref{rho-Stil}), one obtains 
\beq
\sigma(\kappa) = |{\wS}^{-1}(\kappa + i\epsilon)|^{2} \, \rho(\kappa). 
\eeq
The inverse relationship is obtained a follows: since $\left[ {\wS}^{-1}(z)\right]^{-1} = {\wS}(z)$,
one can write using (\ref{sig-S-1til})
\bea
\rho(\kappa) &=& \frac{i}{2\pi} \left[ {\wS}^{-1}(\kappa+i\epsilon)\right]^{-1} 
- \left[ {\wS}^{-1}(\kappa-i\epsilon)\right]^{-1} = R(M_p) \, \delta(\kappa - M_p )  
+ \overline{\rho}(\kappa),
\label{rho-kappa}
\eea
where 
\beq
\overline{\rho}(\kappa) = |{\wS}^{-1}(\kappa + i\epsilon)|^{-2} \, \sigma(\kappa).
\label{rho-bar}
\eeq
In (\ref{rho-kappa}), $M_p$ is a mass pole and $R(M_p)$ the corresponding residue. This pole 
is found as a zero of ${\wS}^{-1}(z)$; when there is more than one pole, one has to sum 
over all poles, when none exit, $R(M_p) = 0$. 

We note that a pole mass is a well-defined concept in QED~\cite{Itzykson:1980rh}, with $M_{\rm p}$ being 
taken as the definition of the electron mass. In QCD, $M_{\rm p}$ can be defined unambiguously 
in perturbation theory only, in which case it is infrared finite and gauge independent to all orders in 
perturbation theory~\cite{Kronfeld:1998di}. Away from perturbation theory, a quark pole mass is thought 
be in conflict with the hypothesis of quark confinement. However, the concept of a pole mass can be of 
phenomenological interest and not necessarily in conflict with confinement: in addition to a zero on the 
real axis, i.e. for $z = M_{\rm p}$ that is real, there might exist complex-conjugate zeros, 
$z = {m_R \pm i m_I}$ and positivity violation. We come back to this discussion in 
section~\ref{sec:model}.

\subsection{Renormalization}
\label{sub:renorm}

Renomalization conditions need to be imposed to determine $Z_\psi(\mu)$ and $Z_m(\mu)$.
Here it is important to make explicit the $\mu$ dependence in the renormalized
quantities. In QCD, for the reasons just discussed, one imposes that at 
some value of an Euclidean momentum $p^2 = - \mu^2 < 0$ the propagator is given by
\beq
S^{-1}(p,\mu) \xrightarrow{p^2 = - \mu^2} \sla{p} - m(\mu),  
\label{ren-cond}
\eeq
where $m(\mu)$ is the renormalized current quark mass. Using the projection operators,
one can easily show that $Z_\psi(\mu)$ and $Z_m(\mu)$ are given in terms of the 
spectral function $\sigma(\kappa,\mu)$ by the following expressions
\beq
Z_\psi(\mu) = 1 - \int^{+\infty}_{-\infty} d\kappa \, \frac{\sigma(\kappa,\mu)}{\kappa^2 + \mu^2},
\label{Zpsi}
\eeq
and 
\beq
Z_\psi(\mu) Z_m(\mu) m(\mu) = m(\mu) + \int^{+\infty}_{-\infty} d\kappa \, \frac{\kappa \sigma(\kappa,\mu)}
{\kappa^2 + \mu^2}.
\label{Zm}
\eeq
Replacing these results in (\ref{Sig-ren}), one obtains for ${\wS}^{-1}(z,\mu)$:
\beq
{\wS}^{-1}(z,\mu) = z - m(\mu) - (z^2 + \mu^2) \int^{+\infty}_{-\infty} d\kappa \, 
\frac{\sigma(\kappa,\mu)}{(z - \kappa)(\kappa^2+\mu^2)}.
\label{S-ren-fin}
\eeq

Notice that (\ref{Z-rho}) and (\ref{Zpsi}) taken together lead to a relationship
between the renormalized spectral function of the propagator, $\rho(\kappa,\mu)$,
and of the its inverse, $\sigma(\kappa,\mu)$, namely: 
\beq
1 - \int^{+\infty}_{-\infty} d\kappa \, \frac{\sigma(\kappa,\mu)}{\kappa^2 + \mu^2}
= \left[\int^{+\infty}_{-\infty} d\kappa \,\rho(\kappa,\mu)\right]^{-1}.
\label{sum-rule}
\eeq
This equality plays an important role in the identification of a positivity violation
of the spectral functions. 

The renormalized $A(p^2,\mu)$ and $B(p^2,\mu)$ functions defined in (\ref{S-AB}) can be written in terms 
of ${\wS}^{-1}(z)$ and using the renormalized expression (\ref{S-ren-fin}) for ${\wS}^{-1}(z)$, they
are given by
\bea
A(p^2,\mu) &=& \frac{1}{2w(p)}\left[{\wS}^{-1}(w(p)+i\epsilon,\mu) - {\wS}^{-1}(-w(p)-i\epsilon,\mu)\right] \nn 
\\
&=& 1 - [p^2 + m^2(\mu)] \int^{+\infty}_{-\infty} d\kappa \, \frac{\sigma(\kappa,\mu)}{(\kappa^2+\mu^2)
[(w(p)+i\epsilon)^2) - \kappa^2]} ,
\label{A-sigma}
\\[0.3true cm]
B(p^2,\mu) &=& - \frac{1}{2}\left[{\wS}^{-1}(w(p)+i\epsilon) + {\wS}^{-1}(-w(p)-i\epsilon)\right] \nn \\
&=& m(\mu) + [p^2 + m^2(\mu)] \int^{+\infty}_{-\infty} d\kappa \, \frac{\sigma(\kappa,\mu) \kappa}
{(\kappa^2+\mu^2)[(w(p)+i\epsilon)^2) - \kappa^2]} .
\label{B-sigma}
\eea

Another renormalization condition is to impose that the propagator has a pole at some
timelike momentum $p^2 = \mu^2 \equiv M^2_{\rm p} > 0$, which is known as the on-shell (os) 
renormalization condition. For completeness, we present the corresponding expressions for the 
renormalization constants and the DSE in this scheme~\cite{{Brown:1968wi},{Krein:1993jb}}:
\bea
&& Z^{\rm os}_\psi(M_{\rm p}) = 1 - \int^{+\infty}_{-\infty} d\kappa \, 
\frac{\sigma(\kappa,M_p)}{(M_{\rm p} - \kappa)^2},
\label{Zpsi-on} \\
&& Z^{\rm os}_\psi(M_{\rm p})[M_{\rm p} - Z^{\rm os}_m \, m(M_{\rm p})] = \int^{+\infty}_{-\infty} d\kappa \, 
\frac{\sigma(\kappa,\mu)}{M_{\rm p} - \kappa},
\label{Zm-on} \\
&& {\wS}^{-1}_{\rm os}(z,M_p) = (z - M_p)\left[1 - (z - M_p) 
\int^{+\infty}_{-\infty} d\kappa \, \frac{\sigma(\kappa,M_p)}{(z - \kappa)(\kappa - M_p)^2}\right].
\label{S-1on}
\eea
The residue of the pole is set to unity.

\section{Model calculation}
\label{sec:model}

We illustrate the application of the formalism with a schematic model for the quark-gluon kernel 
in the Dyson-Schwinger equation (DSE) for quark propagator. We recall that the DSE in its 
unrenormalized form can be written as
\beq
S^{-1}_\Lambda(p) = \sla{p}  - m_\Lambda - i \int \frac{d^4q}{(2\pi)^4} g^2_\Lambda 
\gamma_\mu D^{\mu\nu}_\Lambda(q)
S_{\Lambda}(p-q) T^a \Gamma^a_{\Lambda \, \nu}(q,p-q,p),  
\label{DSE}
\eeq
where $D^{\mu\nu}_\Lambda$ and $\Gamma^a_{\Lambda \, \nu}$ are the unrenormalized gluon propagator 
and quark-gluon vertex function, and $T^a = \lambda^a/2$, $a=1,\cdots,8$, where $\lambda^a$ are 
the color-SU(3) Gell-Mann matrices. 

The model consists in taking for the quark-gluon kernel, $D^{\mu\nu}_\Lambda(p-q)
\Gamma^a_{\Lambda \, \nu}(q,p)$ the following parametrization
\beq
g^2_\Lambda D^{\mu\nu}_\Lambda(q) \Gamma^a_{\Lambda \, \nu}(q,p-q,p) = - g^2 \, 
T^a \, F(q,p-q,p) \, \gamma^\mu,
\label{model}
\eeq
with
\beq
F(q,p-q,p) = \frac{R(q,p-q,p)}{q^2 - \varsigma^2  + i \epsilon},
\label{F}
\eeq
where $\varsigma$ is a mass-scale and $R(q,p-q,p)$ is a singularity-free form-factor that will be
specified shortly ahead. The motivation for this choice is that one expects singularities at 
timelike momentain in the quark-gluon kernel, coming either from the gluon propagator or from the 
quark-gluon vertex, or from both. Certainly, a much more complex singularity structure than 
of a single pole can be expected~\cite{Cornwall:1981zr,Aguilar:2015bud,Bermudez:2017bpx,Aguilar:2018csq}, 
but this single-pole Ansatz serves our needs here for an application to a concrete case. It is also 
rich enough to highlight interesting features of the propagator in Minkowski space. Moreover, it also
serves to emphasize that the appearance of alterations in the analytic properties of 
the propagator can appear in models with no clear connection to QCD. It should be clear that 
with this model, with no proper $\mu-$running of the parameters of the quark-gluon kernel, the mass 
function will be scale dependent. 

Projecting out the Dirac-matrix structure and using the renormalization constants and the 
spectral representation for $S$ under the integral in (\ref{DSE}), the renormalized form of the DSE
can be written as
\beq
{\wS}^{-1}(w(p)+i\epsilon) = Z_{\psi}(\mu) \left[ w(p)  - Z_m(\mu) \,m(\mu) \right]  
+ C_{F} \left(\frac{g}{4\pi}\right)^2 \int^{+\infty}_{-\infty} d\kappa \; K(w(p),\kappa) \; \rho(\kappa,\mu),
\label{K}
\eeq 
where $C_F = T^a T^a = 3/4$, and $K(w(p),\kappa)$ is given by
\beq 
K(w(p),\kappa) = \frac{2}{w(p)}\,  \frac{i}{\pi^2} \int d^4q\, \left[\frac{2 w(p) \kappa - p \cdot (p-q)}
{(p-q)^2  - \kappa^2 + i\epsilon}\right] \frac{R(q,p-q,p)}{q^2 - \varsigma^2 + i \epsilon}.
\eeq
From (\ref{sig-S-1til}), one obtains for $\sigma(\kappa,\mu)$:
\bea
\sigma(\kappa,\mu) &=& C_{F} \left(\frac{g}{4\pi}\right)^2 \int^{+\infty}_{-\infty} d\kappa' \; \frac{1}{2\pi i} 
\left[K(\kappa,\kappa') - K^*(\kappa,\kappa') \right]\; \rho(\kappa',\mu) \nn \\[0.3true cm]
&=& \frac{\alpha_s}{\pi} \, 
\frac{1}{3} \int^{+\infty}_{-\infty} \frac{d\kappa'}{|\kappa|^3}\, \left[\left(\kappa^2-{\kappa'}^2\right)^2 
- \left(\kappa^2 +{\kappa'}^2\right)
+ \varsigma^4 \right]^{1/2} \, \left[(\kappa - \kappa')^2 - 2 \kappa \kappa' - \varsigma^2 \right] 
\nn \\[0.3true cm]
&& \times \, \theta(\kappa^2 - (|\kappa'| + \varsigma)^2) \, R(\varsigma,\kappa',\kappa) \; \rho(\kappa',\mu),
\label{sigma-mod}
\eea
where $\alpha_s = g^2/4\pi$, and $\theta$ is the Heaviside step function.

The problem is solved by iteration: (1) we start with an Ansatz for $\rho(\kappa)$ and
obtain an expression for $\sigma(\kappa)$ from (\ref{sigma-mod}); (2) this $\sigma(\kappa)$ is
inserted in (\ref{S-ren-fin}) and a new $\rho(\kappa)$ is determined from (\ref{rho-kappa}),
where $M_p$ and $R(M_p)$ are determined from ${\wS}^{-1}(z) = 0$; (3) then this 
$\rho(\kappa)$ is used in (\ref{sigma-mod}) to obtain a new $\sigma(\kappa)$. Steps~(1)-(3) are 
repeated until convergence is achieved within a prescribed precision. When using an 
on-shell renormalization, ones uses (\ref{S-1on}) instead of (\ref{S-ren-fin}) in step~(2) of
the iteration and there is no need to determine $M_p$ and $R(M_p)$ because they are set by the 
renormalization condition. 

In the following we present a selected set of results for the spectral functions $\rho(\kappa)$ 
and $\sigma(\kappa)$, and the $A(p^2)$, $B(p^2)$ and $M(p^2) = B(p^2)/A(p^2)$ functions. In this 
toy-model calculation, we use for $R(q,p-q,p) = f(q) f(p-q) f(p)$, with $f(p) = \exp(-|p^2|/\omega^2)$, 
where $\omega$ is a parameter. We use the same $f$ for each leg of the quark-gluon kernel to avoid 
proliferation of parameters. A typical set of parameters, which gives a mass function such that 
$M(p^2 =0) \approx M_p \approx 0.35$~GeV, which is a typical value of a Nambu--Jona-Lasinio 
constituent-quark mass, is the following:
\beq
\mu = 100~{\rm GeV},       \hspace{0.5cm} 
m(\mu) = 0.005~{\rm GeV},   \hspace{0.5cm} 
\alpha_s/\pi = 1.25,       \hspace{0.5cm} 
\varsigma = 0.6~{\rm GeV}, \hspace{0.5cm}
\omega = 2.5~{\rm GeV} .
\label{parms}
\eeq
The choice for the value of $\varsigma$ is inspired on the gluon-mass scale found in 
different studies of the gluon propagator in the infrared{\textemdash}see the review in
Ref.~\cite{Aguilar:2015bud}. Other parameter values around those in~(\ref{parms}) lead to 
the same qualitative results.

\begin{figure}[t]
\begin{center}
\includegraphics[width=220pt]{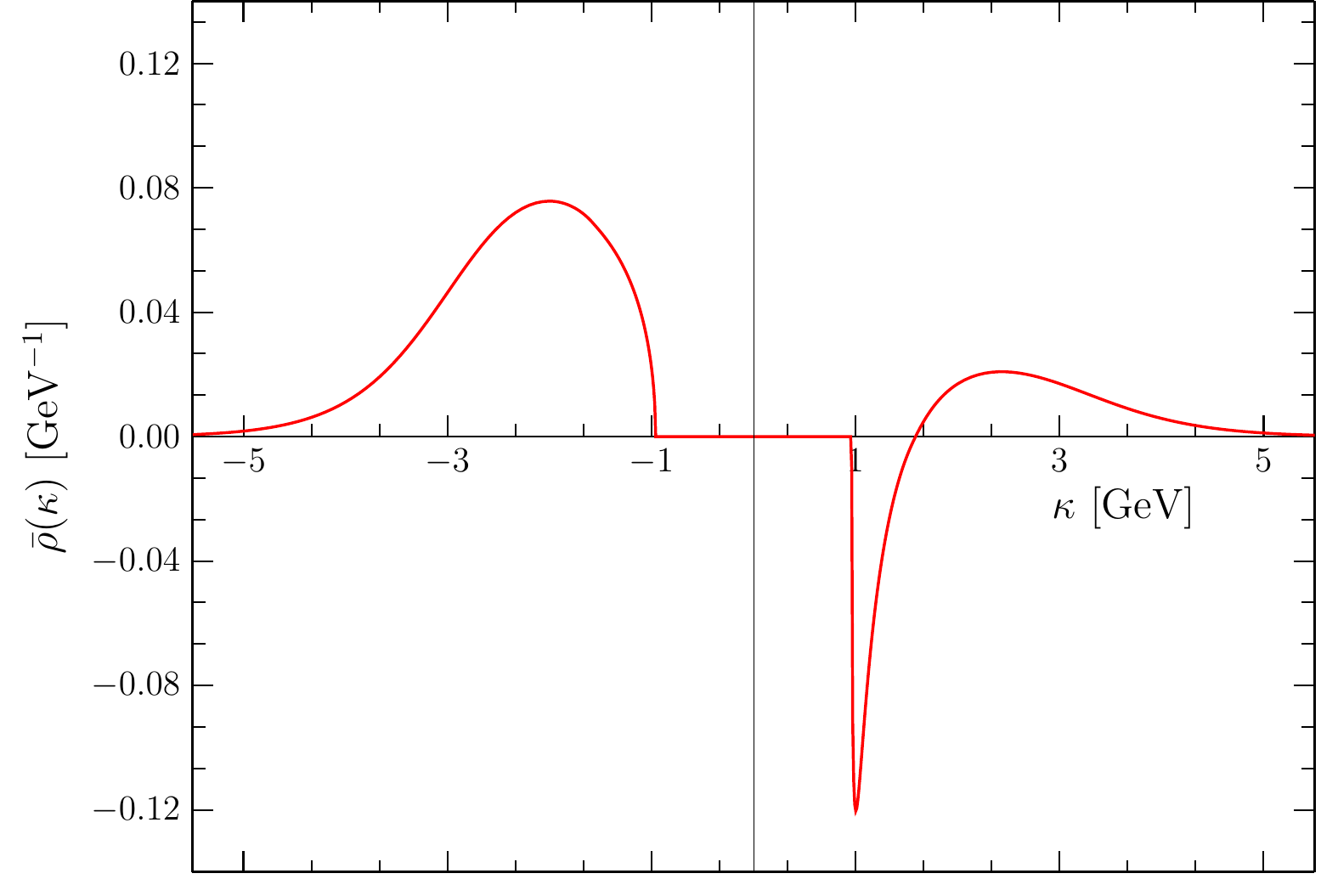}
\hspace{0.15cm}\includegraphics[width=210pt]{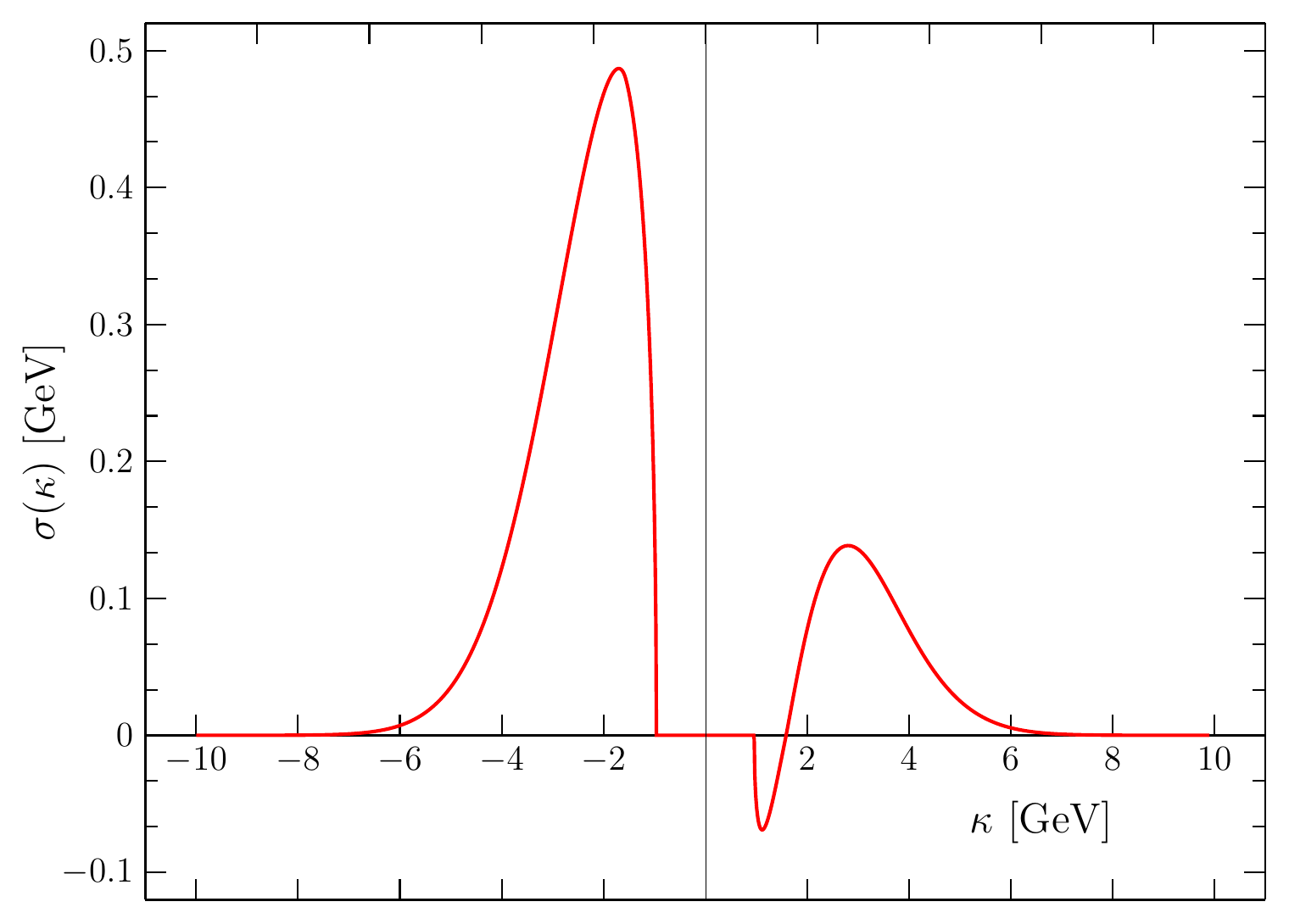}
\end{center}
\caption{The nonpole part $\overline{\rho}(\kappa)$ of the $\rho(\kappa)$ spectral function, 
defined in (\ref{rho-bar}), and the $\sigma(\kappa)$ spectral function.  The parameters 
are those in~(\ref{parms}). 
  }
\label{fig:spect}
\end{figure}

The propagator presents only a real pole, at $M_p = 0.36$~GeV with residue $R(M_p) = 0.83$. 
In Fig.~\ref{fig:spect} we show $\overline{\rho}(\kappa)$ and $\sigma(\kappa)${\textemdash}recall 
that $\overline{\rho}(\kappa)$ is the nonpole part of $\rho(\kappa)$, defined in~(\ref{rho-bar}).
The figure reveals that the spectral functions are negative in a range of~$\kappa$, indicating 
that for this schematic quark-gluon kernel there is positivity violation. We note that positivity 
violation already occurs in an  one-loop calculation of the spectral functions. In an one-loop 
calculation, $\sigma(\kappa)$ is obtained from (\ref{sigma-mod}) by using the delta-function piece 
of $\rho(\kappa)$ in that equation. The positivity violation is not avoided using softer or harder form
factors~$f(q)$. The negative piece in $\sigma(\kappa)$ comes from the 
$-2\kappa\kappa'$ in $\left[(\kappa - \kappa')^2 - 2 \kappa \kappa' - \varsigma^2 \right]$
in (\ref{sigma-mod}). This $-2\kappa\kappa'$ appears because of the $\gamma^\mu$ in the
quark-gluon kernel (\ref{model}); in models with a pseudocalar or a scalar kernel, 
the $-2\kappa\kappa'$ term is absent and the spectral functions are positive~\cite{Brown:1968wi,
Krein:1993jb,Bracco:1993cy,daRocha:1997vk}. Such a positivity violation in the spectral functions
is also seen in the recent investigations of Refs.~\cite{Siringo:2016jrc,Siringo:2017ide} using
other quark-gluon kernels. 

\begin{figure}[t]
\begin{center}
\includegraphics[width=210pt]{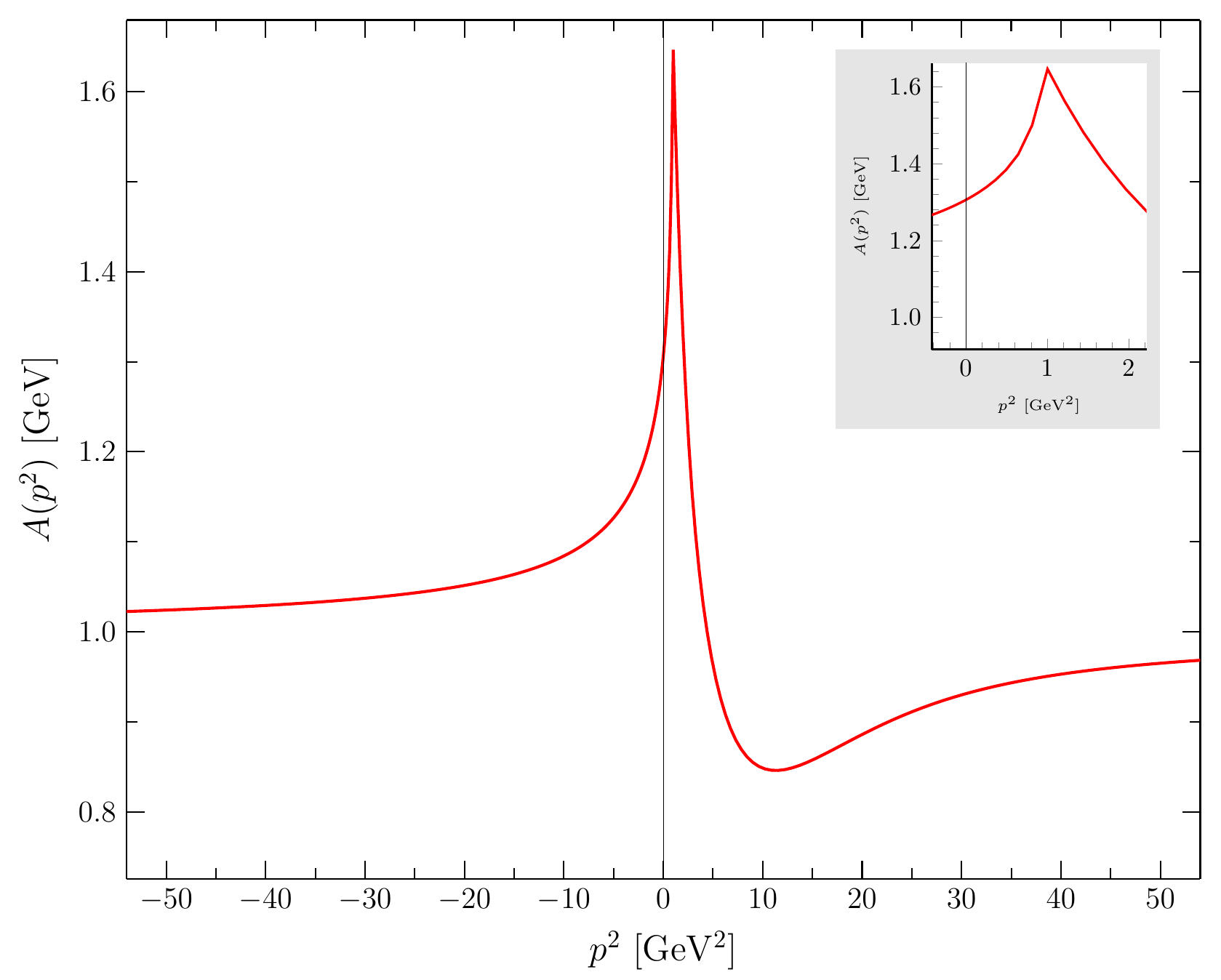}\hspace{0.15cm}
\includegraphics[width=210pt]{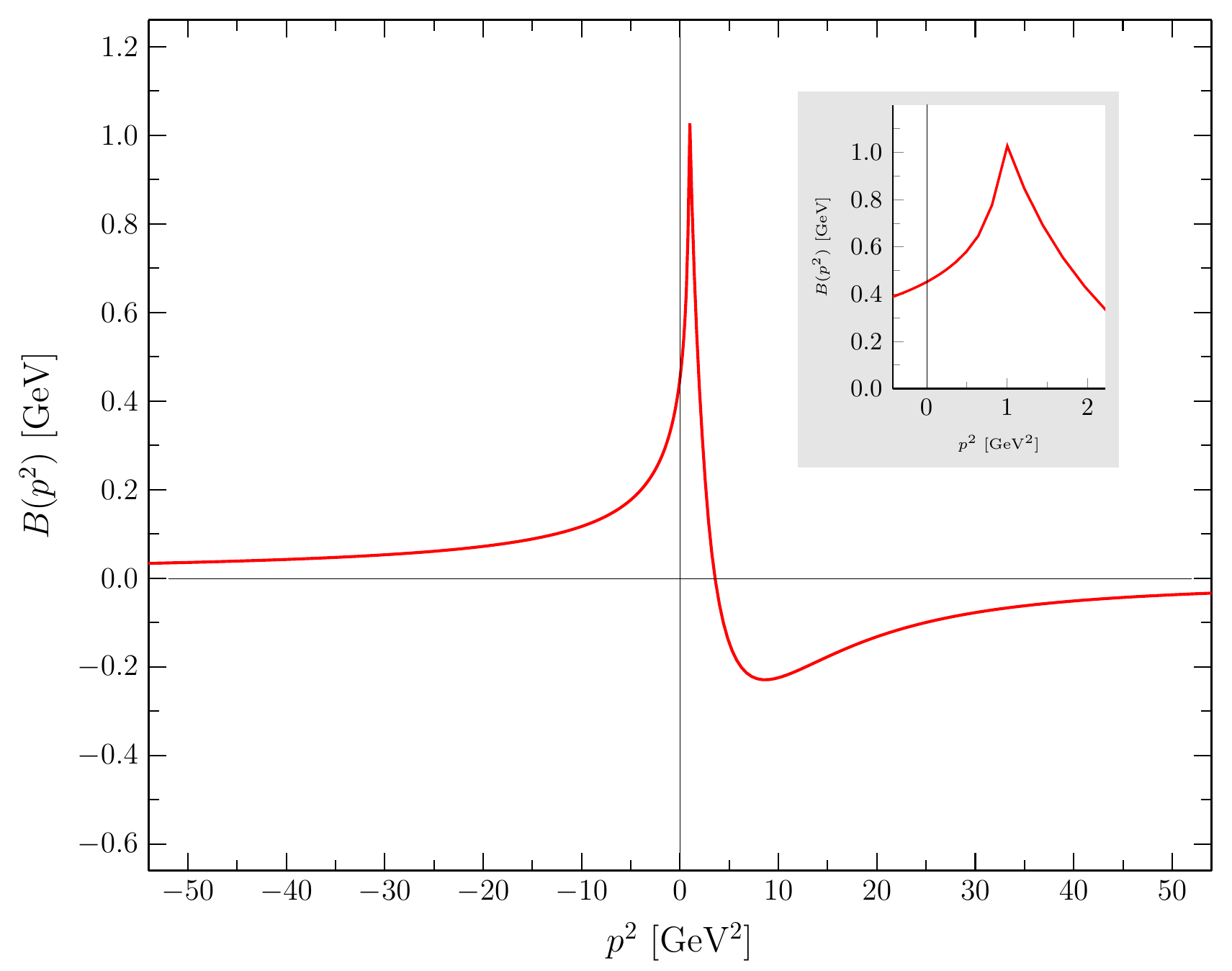}\\[0.5true cm]
\includegraphics[width=210pt]{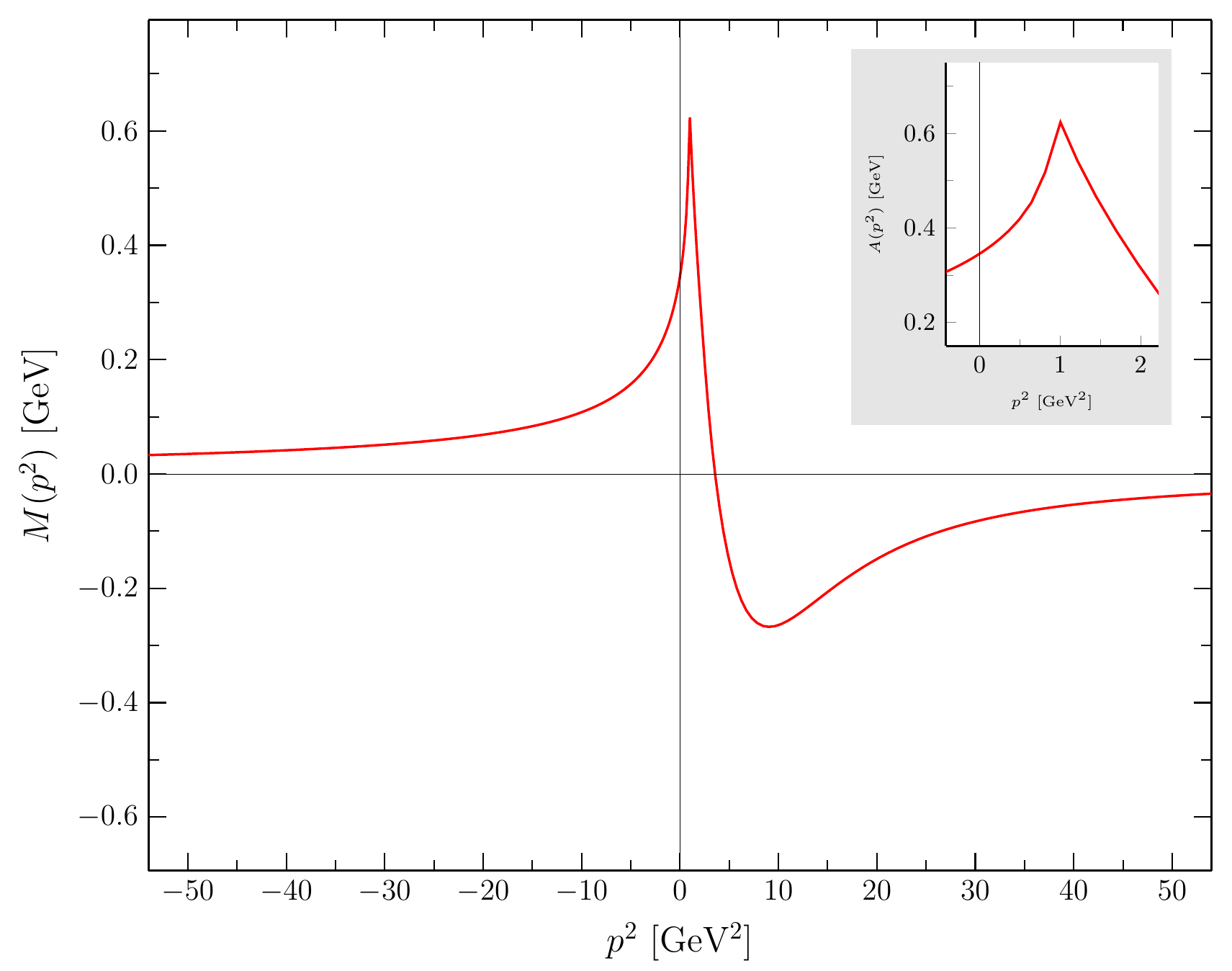}
\end{center}
  \caption{The $A(p^2)$, $B(p^2)$ and $M(p^2)$ functions defined in (\ref{S-AB}). The insets zoom into
  the region around the threshold momentum $p^2 = (M+\varsigma)^2$. Parameters are the same as
  for Fig.~(\ref{fig:spect}).
  }
  \label{fig:ABM}
\end{figure}

Figure~\ref{fig:ABM} presents the solutions for the functions $A(p^2)$, $B(p^2)$ and $M(p^2) 
= B(p^2)/A(p^2)$ defined in~(\ref{S-AB}). The insets in the plots of this figure are zooms into the 
region of the threshold of a branch cut, a cusp at $p^2 = (M_p + \varsigma)^2$. This threshold 
exists because of the pole in~(\ref{F}). The existence of such a cusp, or of more cusps, is a universal 
feature that will appear with any model in which poles exist in the quark-gluon kernel. For example, 
with a ``Landau-gauge'' Ansatz, where the $\gamma^\mu$ in~(\ref{model}) is replaced by 
$(g^{\mu\nu} - q^\mu q^\nu/q^2)\gamma_\nu$, one more cusp appears at $p^2 = M^2_p$,
which obviously comes from the $1/q^2$ in this expression. Such two-cusp structure is clearly 
seen in the corresponding plots in Refs.~\cite{Siringo:2016jrc,Siringo:2017ide} in which the Landau 
gauge is used. Certainly, depending on the complexity of the analytic structure of the quark-gluon 
kernel, one can expect an even more interesting analytic structure in these functions for 
$p^2>0$. 

It is important to note that the positivity violation obtained here {\em is not} related
to the presence of complex-mass poles; for this model-interaction with the parameters  
in~(\ref{parms}), there are no complex-mass poles.
A signal for the presence of such complex-mass poles is the violation of the equality 
in~(\ref{sum-rule}), in that the integral over $\rho(\kappa)$ does not vanish~\cite{Brown:1968wi,
Krein:1993jb}{\textemdash}in the present calculation, the equality is respected. Complex-mass 
poles are found in this model, but for parameters very different from those in~(\ref{parms}). 
Moreover, positivity violation is also obtained when using the on-shell renomalization scheme, 
specified by the equations~(\ref{Zpsi-on})-(\ref{S-1on}). In this scheme, the pole mass is set 
as a renormalization condition. In this scheme, positivity violation also does not depend on the 
values of the parameters employed. It is also obtained in a one-loop calculation in this scheme. 
The origin of the positivity violation is again the term $-2\kappa\kappa'$, discussed above. 
Moreover, complex-mass poles are also found in the on-shell scheme for certain values of the 
parameters. The appearance of such complex-mass poles, in any of the two renormalization schemes,
is closely related to the ultraviolet behavior of the quark-gluon kernel: when the interaction is 
screened in the ultraviolet, the complex-mass poles recede to infinity, in a way very similar to 
the meson-baryon case investigated in Refs.~\cite{Krein:1993jb,Bracco:1993cy,
daRocha:1997vk}. We leave for a future work the investigation of these issues~\cite{Krein:2014zaa}.

%
\section{Conclusions and perspectives}
\label{sec:concl}

We presented a computational method for solving the Dyson-Schwinger equation (DSE) for the 
quark propagator in Minkowski space. The method uses the spectral representation of the 
propagator and of its inverse. Instead of solving the DSE for the momentum-dependent 
mass and renormalization functions, as it is usually solved in its Euclidean formulation,
in the spectral method the DSE is solved for the spectral functions. As an application of 
the formalism, we has used a schematic model for the quark-gluon kernel. Although the model 
is a very crude representation of the singularity structure of the quark-gluon kernel in 
Minkowski space, it served the purpose to illustrate important features of the solutions, 
such as the presence of poles and branch cuts, and positivity violation. 

The model calculation served to illustrate in particular the fact that one can obtain positivity 
violation in the quark propagator in models whose connection to QCD is not a priori clear.
We recall that positivity violation in the propagator is conjectured to be relevant 
for the interpretation of quark confinement, it that it implies that a quark can not
be associated with an asymptotic state. We have shown that positive
violation is not necessarily related to the presence of complex-mass poles, as they are
absent in this model-interaction for the parameters used in the calculation. Moreover, we 
have pointed out that positivity violation is also obtained in an one-loop perturbative 
calculation, and is independent of the renormalization conditions. These results have obvious 
impact for calculations obtaining positivity violation using different models, as it poses the 
question whether it is a real feature of QCD or of the particular model and of the particular 
truncation used to solve the DSE. 

Future work includes examination of these issues in more realistic models of QCD. One envisages
progress with methods based on systematic improvements on a zeroth-order approximation
that captures essential features of QCD, examples of which are those of 
Refs.~\cite{Krein:2014zaa,PhysRevD.92.074027,Siringo:2016jrc,PhysRevD.96.114011}.

\begin{acknowledgements}
Work partially supported by: Coordena\c{c}\~ao de Aperfei\c{c}oamento de Pessoal de N\'{\i}vel Superior - CAPES,
Grants. no 8888.330776 (C.S.R.C.), 8888.330775 (E.L.S.), 8888.330773 (V.V.L.), 
Conselho Nacional de Desenvolvimento Cient\'{\i}fico e Tecnol\'ogico 
- CNPq, Grants No. 305894/2009-9 (G.K.), 464898/2014-5(G.K) (INCT F\'{\i}sica 
Nuclear e Apli\-ca\-\c{c}\~oes), Funda\c{c}\~ao de Amparo \`a 
Pesquisa do Estado de S\~ao Paulo - FAPESP, Grant No. 2013/01907-0 (G.K.). 
\end{acknowledgements}

\bibliographystyle{spphys}       


\bibliography{Faddeev_sclk}

\end{document}